\documentclass[a4paper,12pt]{article} 
\usepackage{amsfonts}

\title{Induced commutative Chern-Simons term from noncommutativity in
  planar systems} 
\author{M.~L.~Ciccolini$^a$~\footnote{Electronic address:
    mciccoli@ph.ed.ac.uk}, C.~D.~Fosco$^b$~\footnote{Electronic
    address: fosco@cab.cnea.gov.ar} and
  A.~L{\'o}pez$^b$~\footnote{Electronic address: lopezana@cab.cnea.gov.ar}
  \\ \\
  {\normalsize\it $^a$Department of Physics and Astronomy, University of Edinburgh,}\\
  {\normalsize\it Edinburgh, EH9 3JZ, Scotland.}\\ \\
  {\normalsize\it $^b$Centro At{\'o}mico Bariloche - Instituto Balseiro,}\\
  {\normalsize\it Comisi{\'o}n Nacional de Energ{\'\i}a At{\'o}mica}\\
  {\normalsize\it 8400 Bariloche, Argentina.}}  
\date{}
\begin{document}
\maketitle
\begin{abstract}
\noindent We test the consistency of the use of a noncommutative 
theory description for charged particles in a strong magnetic field,
by deriving the induced Chern-Simons (CS) term for an external Abelian
gauge field in $2+1$ dimensions.  In this description, the system is
modeled by a noncommutative matter field coupled to a $U(1)$
noncommutative gauge field, related to the original, commutative one,
by a Seiberg-Witten transformation.  We show that an Abelian CS term
for the commutative gauge field is indeed induced, and moreover that
it matches the result of previous commutative field theory
calculations.
\end{abstract}
\newpage
%%%%%%%%%%%%%%%%%%%%%%%%%%%%%%%%%%%%%%%%%%%%%%%%%%%%%%%%%%%%%%%%%%%%%%
%%%%%%%%%%%%%%%%%%%%%%%%%%%%%%%%%%%%%%%%%%%%%%%%%%%%%%%%%%%%%%%%%%%%%%
%%%%%%%%%%%%%%%%%%%%%%%%%%%% Introduction %%%%%%%%%%%%%%%%%%%%%%%%%%%%
%%%%%%%%%%%%%%%%%%%%%%%%%%%%%%%%%%%%%%%%%%%%%%%%%%%%%%%%%%%%%%%%%%%%%%
%%%%%%%%%%%%%%%%%%%%%%%%%%%%%%%%%%%%%%%%%%%%%%%%%%%%%%%%%%%%%%%%%%%%%%
\section{Introduction}\label{sec:intro}
The relevance of noncommutative field theories for the description of
2+1 dimensional systems in a strong external magnetic field, notably
the Quantum Hall effect, hardly needs to be stressed.  Indeed, it has
been shown some time ago that noncommutative field theories in $2+1$
dimensions may be a useful tool for the approximate description of
some planar condensed matter models, when the kinetic term may be
neglected in comparison with the coupling between the matter current
and the external field. When this is the case, the two spatial
coordinates become conjugate to each other, so that the system may be
thought of as defined on a `quantum phase space': a noncommutative
quantum field theory \cite{susskind,PP,KS,MS,PA,LMR,DN,AP,cesarana}.  
In such
 an effective description, the noncommutativity of space is
characterized by the antisymmetric (`Poisson') tensor $\theta^{ij}$:
\begin{equation}
\theta^{ij}\;=\; \theta \, \epsilon^{ij} \;\;\;,\;\;\; 
\theta \;=\; - (eB)^{-1} \;\;\;\; i,j \,=\, 1,2.
\end{equation}
Hence, all the dependence on the external magnetic field is traded for
the noncommutativity, so that if $B$ manifests itself through, for
example, parity or time-reversal symmetry breaking effects, this has
to appear as a pure consequence of the noncommutativity of the
effective description.

A particularly interesting opportunity to check the noncommutative
description arises when one considers a commutative theory
corresponding to charged particles in the presence of a strong
magnetic field and of an abelian gauge field ${\mathcal A}_\mu$
(unrelated to $B$) which plays the role of a `source' or external
probe. It is well known that, in a purely commutative description, one
may apply perturbation theory to obtain the induced CS term for
${\mathcal A}_\mu$. Of course this is only possible in the case in which
the relation between the external magnetic field ($B$) and the
electronic density ($\rho$) is such that there is an integer number of
filled Landau levels, i.e., the filling fraction is \mbox{$\nu\,=\,
 { \frac{\rho }{e B}} 2\pi\,=\,n$} with $n$ integer (in units in 
which $\hbar=c=1$).
In this evaluation, the constant magnetic field is taken into account
exactly, since the full propagator in the presence of the magnetic
field and at a finite density is used in the perturbative
series~\cite{lf}. The emergence of an induced CS term may be seen, in
this picture, to be a parity breaking effect due to the magnetic
field.  In the more general case, i.e., when a perturbative
calculation is not correct, it can be argued, based on symmetry
reasons, that there should be a Chern-Simons term in the effective
action for the external probe. Equivalently, due to the presence of
the external magnetic field there should be a transverse response to a
static electric field. In two spatial dimensions, such a response can
be obtained only if there is a Chern-Simons term. Moreover, the
coefficient should be determined by the Hall conductance which can be
shown to be $\sigma_{xy}={\frac {\rho c e}{B}}$ in a translational invariant
system~\cite{PG}.

Our aim, in this paper, is to show that the same object, namely, the
commutative induced CS term, may also be derived in the corresponding
effective noncommutative framework. This result will look, at first
sight, paradoxical, since a perturbative evaluation of the effective
action in the noncommutative model yields no induced CS term for the
{\em noncommutative\/} gauge field. 

The organization of this paper is as follows: in
section~\ref{sec:eff}, we derive the noncommutative theory as an
effective description of the original commutative model, in
section~\ref{sec:ics} we present a calculation of the induced CS term
in the noncommutative theory context, and in section~\ref{sec:conc} we
present our conclusions.

\section{The noncommutative effective theory}\label{sec:eff}
To begin with, let us assume that the commutative system is defined by
a generating functional ${\mathcal Z}_B[{\mathcal A}_\mu]$, where
${\mathcal A}_\mu$ denotes an external source
\begin{equation}
  \label{eq:zcomm}
  {\mathcal Z}_B[{\mathcal A}_\mu]\;=\; \int { \mathcal D}{\bar\psi}\,
{\mathcal D}\psi \, \exp \left\{i\, S[{\bar\psi},\psi;B,{\mathcal A}_\mu] \right\}     \;,
\end{equation}
and we have made explicit the fact that the action $S$ depends on
the magnetic field $B$. The generating functional has been written 
as a path integral over a charged field, which will typically be a 
spinless fermionic field. The corresponding action for this 
(non-relativistic, Grassmann) field is 
\begin{equation}
S[{\bar \psi},\psi;B, {\mathcal A}] \;=\; \int d^3x \left[ {\bar\psi} 
i  D_0 \psi - \frac{1}{2m} \overline{D_i \psi} (D_i \psi) \right],
\end{equation}
where the covariant derivatives are defined by
\begin{equation}
  \label{eq:defcov}
D_0 \,=\, \partial_0 - ie(a_0 + {\mathcal A}_0)
\;\;\;\;\;\;
D_i \,=\, \partial_i - ie(A_i + {\mathcal A}_i)\;,
\end{equation}
where $a_0$ is a constant, which plays the role of a chemical
potential, fixing the total charge $Q$ of the system, and $A_i$
($i=1,2$) is the vector potential of the constant magnetic field,
namely: $\epsilon^{ij} \partial_i A_j = B$. In order to understand the emergence of
an effective noncommutative description, we shall first set ${\mathcal
  A}_\mu = 0$, and then reintroduce it (as well as possible
interactions) afterwards, so that the generating functional we will
first consider is
\begin{equation} 
{\mathcal Z}_B\,\equiv\,{\mathcal Z}_B[0]\;=\; \int{\mathcal D}{\bar\psi}{\mathcal D}\psi \;
\exp \left \{ i S[{\bar \psi},\psi;B,0] \right\}\;.
\end{equation}
We then use the complex combinations $D = D_1 + i D_2$ and 
${\overline D} = D_1 - i D_2$, plus an integration by parts, to write
$$
S[{\bar \psi},\psi;B,0]\;=\;\int d^3x \left[ {\bar\psi} \left(i \partial_t + e a_0 - 
\frac{ e B}{2 m }\right)\psi \right.
$$
\begin{equation}
\left. + \,\frac{1}{2m} {\bar\psi} {\overline D} D \psi \right],
\end{equation}
where we have used the elementary relation 
$[D_1, D_2] = -i e B$.
The noncommutative theory will be obtained by performing an 
approximation which relies upon the smallness of the scale defined 
by the mass in comparison with the one defined by the magnetic 
field $B$. Moreover, in the noncommutative theory, the matter 
fields will verify first order equations. In order to take 
these facts into account, it is convenient to use an equivalent 
action  ${\widetilde S}$, obtained by introducing two auxiliary 
complex fields $\lambda$ and ${\bar\lambda}$:
$$
{\widetilde S}[{\bar \psi},\psi,{\bar \lambda},\lambda;B]\;=\;\int d^3x \left[ 
{\bar\psi} \left(i  \partial_t + e a_0 - \frac{ e B}{2 m }\right)\psi \right.
$$
\begin{equation}
 \label{eq:hst}
\left. + {\bar\lambda}D \psi \,+\,  {\bar\psi}{\bar D} \lambda \,-\,
\frac{m}{2}{\bar\lambda}\lambda \right]\,,
\end{equation}
which, of course, reproduces $S$ when ${\bar\lambda}$ and $\lambda$ are 
integrated out.

Assuming the charge density of the system to be fixed to some value
$\rho$, we may write $e a_0=-\frac{2\pi }{m} \rho$, thus we see that the $e
a_0$ term scales like $m^{-1}$.  In fact, all the terms in the first
line of (\ref{eq:hst}) have the same dependence in $m$, since when one
uses an expansion of the time dependence of the fields in terms of the
Landau level energies
\begin{equation}
E_n \;=\;  \omega_c \left(n+\frac{1}{2}\right) \;\;\;,\;\; \omega_c \,=\, \frac{|e B|}{m }\;\;, 
\end{equation}
the time derivative also picks up a factor of $m^{-1}$, regardless of
the level considered. The noncommutative description arises when the
$m \to 0$ limit is taken.  This requires the vanishing of the term 
which
is quadratic in the Grassmann fields. Using the explicit values of the
energies, we see that this implies a constraint between the density
and the magnetic field:
\begin{equation}
  \label{eq:const}
\rho \,=\, |e B|\frac{n}{ 2\pi}  \;\;\; n \in {\mathbb N} \;. 
\end{equation}  
In this limit, the term quadratic in the auxiliary field vanishes, and
the equations of motion are first order: $D \psi = 0$, i.e., the fields
are in the lowest Landau level (LLL). This means that only the $n=1$
case actually appears in (\ref{eq:const}), when $m$ is strictly zero.
This reduction to the LLL implies that, if there are more terms in the
action (like interactions) one has to replace the standard product by
the Moyal product: the theory becomes noncommutative~\cite{cesarana}.
This is true, in particular, for the terms that correspond to the
coupling to the external field ${\mathcal A}_\mu$. We start from that
standpoint in the following section.
%%%%%%%%%%%%%%%%%%%%%%%%%%%%%%%%%%%%%%%%%%%%%%%%%%%%%%%%%%%%%%%%%%%%%%%%%%%%%%
%%%%%%%%%%%%%%%%%%%%%%%%%%%%%%%%%%%%%%%%%%%%%%%%%%%%%%%%%%%%%%%%%%%%%%%%%%%%%%
\section{Induced Chern-Simons term}\label{sec:ics}
The existence of an effective noncommutative description when the
fields are projected to the LLL amounts to:
\begin{equation}
  \label{eq:zs}
  {\mathcal Z}_B[{\mathcal A}_\mu] \;\simeq\; 
\widehat{\mathcal Z}_\theta[{\widehat {\mathcal A}}^\theta_\mu] \;\;\;(m \to 0)  
\end{equation}
where $\widehat{\mathcal Z}_\theta[{\widehat {\mathcal A}}^\theta_\mu]$ is the
generating functional for the noncommutative theory 
\begin{equation}
  \label{eq:zncomm}
  \widehat{\mathcal Z}_\theta[{\widehat {\mathcal A}}^\theta_\mu]\;=\; 
\int {\mathcal D}{\widehat {\bar \psi}}
{\mathcal D}{\widehat \psi} \, \exp \{i S^\theta[{\widehat{\bar \psi}},
{\widehat \psi};0,{\widehat {\mathcal A}}^\theta_\mu] \}     \;.
\end{equation}
Independently of the details of the particular model considered, the
noncommutative action will be assumed to have the same structure as
the commutative one, except for the fact that there will be no
coupling to an external (noncommutative) magnetic field ($B$ is
entirely absorbed in $\theta$).  As we mentioned in
section~\ref{sec:intro}, when there is an integer number of filled
Landau levels, a local CS term in ${\mathcal A}_\mu$ appears as the
leading term in a derivative expansion of the commutative generating
functional, namely,
\begin{equation}
  \label{eq:zcexp}
  {\mathcal Z}_B[{\mathcal A}_\mu]\;\simeq\;\exp \{i \kappa(B) \, S_{CS}[{\mathcal A}_\mu] \} \;,
\end{equation} 
where $S_{CS}$ denotes the CS action,
\begin{equation}
  \label{eq:defscs}
S_{CS}[A]\;=\;\frac{1}{2}\int d^3x\,\epsilon^{\mu\nu\lambda} A_\mu\partial_\nu A_\lambda  
\end{equation}
while $\kappa$ is the corresponding coefficient, determined by the number
of filled Landau levels $n$, through the relation:
\begin{equation}
  \label{eq:defk}
\kappa(B) \;=\;{\frac {n}{2\pi}}\;=\;{\frac { \rho e}{B}} \;.
\end{equation}
Our aim is to show that the same object can be derived in the
noncommutative theory framework.

We shall provide an explicit calculation. To that end, we will assume
that the commutative theory action $S$ is given by:
\begin{equation}
S\;=\;\int d^3x \left[ {\bar \psi} i D_0 \psi -\frac{1}{2m}( \overline{D_i \psi})  (D_i \psi) 
\right],
\end{equation}
where $D_j \;=\;\partial_j + A^B_j + A_j$, with $A^B_j$ denoting the part of
the gauge field that corresponds to the magnetic field $B$, i.e.,
$\epsilon^{jk}\partial_j A_k = i e B$, and \mbox{$D_0 \;=\;\partial_0 + i \mu + A_0$}, 
where $\mu$ is the chemical potential, to be fixed latter, in terms of 
the total charge of the system.  To simplify the comparison with the
noncommutative theory, we shall use the convention that $A_\mu$ is
anti-Hermitian, and moreover that the charge $e$ has been absorbed 
into the gauge field definition.
%%%%%%%%%%%%%%%%%%%%%%%%%%%%%%%%%%%%%%%%%%%%%%%%%%%%%%%%%%%%%%%%%%%%%
%%%%%%%%%%%%%%%%%%%%%%%%%%%%%%%%%%%%%%%%%%%%%%%%%%%%%%%%%%%%%%%%%%%%%
%%%%%%%%%%%%%%%%%%%%%%% Noncommutative description %%%%%%%%%%%%%%%%%%
%%%%%%%%%%%%%%%%%%%%%%%%%%%%%%%%%%%%%%%%%%%%%%%%%%%%%%%%%%%%%%%%%%%%%
%%%%%%%%%%%%%%%%%%%%%%%%%%%%%%%%%%%%%%%%%%%%%%%%%%%%%%%%%%%%%%%%%%%%%
The noncommutative action $S^\theta$ is, on the other hand, explicitly
given by
\begin{equation}
S^\theta\;=\;\int d^3 x \left[ {i \bar {\widehat{\psi}}} \star {\widehat D}_0 
{\widehat \psi} \,-\, 
\frac{1}{2m}  (\overline {{\widehat D}_i {\widehat  \psi}}) \star {\widehat D}_i 
{\widehat \psi} \right]
\end{equation}
where $\star$ denotes the noncommutative Moyal product:
\begin{equation}
\left.  f(x)\star g(x) \;=\; \exp \left( -\frac{i}{2}\theta \epsilon^{ij}
\frac{\partial}{\partial \zeta^i} \frac{\partial}{\partial \xi^j} \right)
f(x+\zeta)g(x+\xi) \right|_{\zeta=0,\xi=0} \;,
\end{equation}
${\widehat D}_j {\widehat \psi} = \partial_j {\widehat \psi} + {\widehat \psi} \star
{\widehat A}_j$, and ${\widehat D}_0 = \partial_0 {\widehat \psi} + {\widehat \psi}
\star ({\widehat A}_0 + i \mu)$ (we have chosen the antifundamental
representation).
After some standard manipulations, we may rewrite $S^\theta$ as:
\begin{equation}
S^\theta\;=\;S_{\mathrm{free}}\,+\,S_{\mathrm{int}}
\end{equation}
where
\begin{eqnarray}
S_{\mathrm{free}} & = & \int d^3x \left[ i{\overline {\widehat \psi}} \star( \partial_0 +i \mu) 
{\widehat \psi} - 
\frac{1}{2m} \partial_i{\overline {\widehat \psi}} \star \partial_i{\widehat \psi} 
\right] \nonumber\\
S_{\mathrm{int}} & = & \int d^3x \left\{i {\widehat \psi} \star {\widehat A}_0 \star 
{\overline {\widehat\psi}} -\frac{1}{2m} \left[ {\widehat \psi} \star {\widehat A_i} 
\star \partial_i  {\overline {\widehat\psi}} - \partial_i {\widehat \psi} 
\star {\widehat A}_i \star  {\overline {\widehat\psi}}
\right. \right. \nonumber \\ & & \left. \left .
-{\widehat \psi} \star {\widehat A}_i \star {\widehat A}_i 
\star {\overline {\widehat\psi}} \right] \right\} \;.
\end{eqnarray}

The induced CS term will be obtained, as usual, from the fermionic effective action 
$S_{\mathrm{eff}}$,
\begin{equation}
\widehat{\mathcal{Z}}^\theta [{\widehat A}^\theta_\mu]\;=\; 
\int \mathcal{D}{\overline {\widehat\psi}} \mathcal{D}{\widehat \psi}
\exp \left[ i(S_\mathrm{free}+S_\mathrm{int}) \right] = 
\exp \left( iS_\mathrm{eff}[{\widehat A}] \right)\;,
\end{equation}
with:
%\begin{eqnarray}
%S_{\mathrm{eff}} & = & \mathrm{Tr} \left\{ \ln \left[
% k_0-\frac{\vec{k}^2}{2m} - \mu 
% + i \int dp \left(  \tilde A_0(p)  + \frac{(2\vec{k}+\vec{p}) \vec{
% \tilde A} (p)} {2m} \right)
%e^{-i/2 \; k\theta p}
%\right.\right. \nonumber \\
%&  &
%\left.\left.
%+\int \frac{dp\;dq}{2m} \vec{\tilde A}(p)\vec{\tilde A}(q)
%e^{-i/2 \xi}
%\right]\right\}
%\end{eqnarray}
%%%%%%%%%%%%%%%%%%%%%%
\begin{eqnarray}
S_{\mathrm{eff}} & = & \mathrm{Tr} \left\{ \ln \left[
 (k_0-\frac{\vec{k}^2}{2m} - \mu)
 \delta(k_1-k_2) \right. \right. \nonumber \\ 
& &  + i \int dp \left(  \tilde A_0(p)  + \frac{(\vec{k}_1+\vec{k}_2) \vec{
 \tilde A} (p)} {2m} \right)
e^{-i/2 \; k\theta p}
 \delta(k_1+p-k_2)
 \nonumber \\
&  &
\left.\left.
+\int \frac{dp\;dq}{2m} \vec{\tilde A}(p)\vec{\tilde A}(q)
e^{-i/2 \xi}
 \delta(k_1+p+q-k_2)
\right]\right\}
\end{eqnarray}
%%%%%%%%%%%%%%%%%%%%%%
and ${\widehat A}_\mu(x)\,=\,\int d^3p {\widetilde A}_\mu (p) \, e^{i p\cdot x}$. We
are using the following notation:
\begin{eqnarray*}
\xi & = & k_1 \theta p  + k_1 \theta q - k_1\theta k_2
      +p \theta q - p\theta k_2
      -q\theta k_2 \\
p \theta q & =  & p_i \theta ^{ij} q_j = \theta p_i \epsilon ^{ij} q_j .
\end{eqnarray*}
Expanding $S_{\mathrm{eff}}$ up to second order in $A_\mu$, we find:
\begin{equation}
S_{\mathrm{eff}} \;\simeq\;S_{\mathrm{eff}}^{(0)}\,+\, S_{\mathrm{eff}}^{(1)}\,+\, S_{\mathrm{eff}}^{(2)}
\end{equation}
where
\begin{eqnarray}
S_{\mathrm{eff}}^{(0)} & = & \mathrm{Tr} \left[
\ln \left(
\Delta^{-1}(k_1,\mu)\delta(k_1-k_2)
\right) \right]
\\
S_{\mathrm{eff}}^{(1)} & = & \mathrm{Tr} 
\left[
\Delta(k_1,\mu) \delta (k_1-k_2)
T_1(k_2,k_3)
\right]
\\
S_{\mathrm{eff}}^{(2)} & = &
\mathrm{Tr}
\left[
\Delta(k_1,\mu)\delta(k_1-k_2)T_2(k_2,k_3)-\frac{1}{2}\Delta(k_1,\mu)\delta(k_1-k_2)
\right.
\nonumber \\
& & 
\left.
T_1(k_2,k_3)\Delta(k_3,\mu)\delta(k_3-k_4)T_1(k_4,k_5)
\right]
\end{eqnarray}
and  we defined
\begin{eqnarray*}
\Delta(k, \mu) & = & \left( k_0-\frac{\vec{k}^2}{2m} - \mu \right)^{-1} \\
T_1 (k_1,k_2)& = & i\int d^d p \left( 
  \tilde A_0(p)  + \frac{(\vec{k_1}+\vec{k_2}) \vec{
 \tilde A} (p)} {2m} \right) 
\delta(p+k_1-k_2) e^{-i/2\,k_1 \theta p} \\
T_2 (k_1,k_2)& = & \int \frac{dp\;dq}{2m} \vec{\tilde A}(p)\vec{\tilde A}(q)
e^{-i/2 \xi} \delta (p+q+k_1-k_2)\;.
\end{eqnarray*}
By analogy with the commutative field theory calculation, one may be
inclined to think that the CS term comes from the term of second order
in $A_\mu$:
\begin{eqnarray}
S_\mathrm{eff}^{(2)} & = & \int d^3p d^3k 
\Delta(k,\mu) \left\{ \frac{\vec{\tilde A}(-p) \vec{\tilde A}(p)}{2m}
+\frac{1}{2}\Delta(k+p,\mu)\left[\tilde A_0(p) + \frac{2\vec{k}+\vec{p}}{2m} 
\vec{\tilde A}(p)\right]
\right.
\nonumber \\
& &
\label{seff2}
\left.
\left[
\tilde A_0(-p) + \frac{2\vec{k}+\vec{p}}{2m} \vec{\tilde A}(-p)
\right] \exp (i/2 \; p \theta p)
\right\}\;;
\end{eqnarray}
however, it is straightforward to check that the transverse
conductivity coming from that term vanishes. This result has been
derived under the assumption $[\partial_i,\partial_j] A_\mu =0$, namely, `trivial'
$A_\mu$ configurations. At the level of the calculation, this amounts to
using the relation: $\exp(i p\theta p) {\widetilde A}_\mu(p) = {\widetilde
  A}_\mu (p)$ in Fourier space. Relaxing that condition, however, does
not help to induce a CS term, in spite of the fact that it might
produce parity breaking effects for some special, singular
configurations (i.e., vortex like configurations) of the gauge field.

The resolution of this apparent paradox lies in the fact that the
relation between the commutative and noncommutative generating
functionals is less direct than it seems, since the commutative one
depends on $A$ while the noncommutative functional has ${\widehat
  A}^\theta_\mu$ as its argument.  The relation between these two gauge
fields is determined by the requisite that the mapping should preserve
the respective orbits of each theory. It is given by the
Seiberg-Witten map~\cite{seibergwitten}, which provides a relation of
the type:
\begin{equation}
{\widehat A}^\theta_\mu \;=\; {\widehat A}^\theta_\mu(A,\theta)
\end{equation}
where the dependence is highly nonlinear, and in general it may be
found only by applying some sort of approximation scheme. Thus, our
claim is that, in order to recover the induced CS term of the
commutative theory, one should use the SW relation to bring the
effective action back to a functional of $A_\mu$, the commutative gauge
field. Since the CS term is quadratic in $A_\mu$, the use of a first
order expansion in $\theta$ is sufficient, since a $\theta$ expansion of the
Seiberg-Witten map yields~\cite{seibergwitten,bichletal}:
\begin{equation}
\widehat{A}_\mu \;=\; A_\mu \,-\, \frac{\theta^{\rho\sigma}}{2} A_\rho 
\{\partial_\sigma A_\mu + F_{\sigma\mu} \}\,+\,{\mathcal O}(\theta^2) \;.
\end{equation}
With this in mind, we may now calculate the term $S_q[A]$,  of second order 
in the {\em commutative\/} gauge field $A$, coming from the  effective action, 
obtaining:
\begin{equation}
S_q[A]\;=\; S_{\mathrm{even}} + S_{\mathrm{odd}}
\end{equation}
with $S_{\mathrm{even}}$ given by $S_{\mathrm{eff}}^{(2)}$ as in eq.~(\ref{seff2}),
 and
$$
S_{\mathrm{odd}}[A] \;=\; -i \frac{\theta ^ {\rho \sigma}}{2}  
\int d^3x \int \frac{d^3k}{(2\pi)^3} \Delta (k,\mu)\;
A_\rho(x) [ (\partial_\sigma A_0 +F_{\sigma\,0})
$$
\begin{equation}
+ \frac{2\vec{k}}{2m} (\partial_\sigma \vec{A} + \vec{F_\sigma})] 
\;.
\end{equation}
where we have assumed trivial configurations for the field $A_\mu(x)$. 
Finally, we may write the explicit form of the $S_{\mathrm{odd}}$ term:
\begin{equation}
 S_{\mathrm{odd}} \;=\; - i I(\mu) \, \frac{\theta ^ {\rho \sigma}}{2} \, 
\int d^3x \, A_\rho (\partial_\sigma A_0 +F_{\sigma\,0}) 
\end{equation}
where
\begin{equation}
I(\mu)\;=\;\int \frac{d^3k}{(2\pi)^3}\, \Delta (k,\mu) \;=\; -\, \frac{i}{2\pi} \,  m \mu  \;.
\end{equation}
%%%
Using this result, the relation $\theta = - (e B)^{-1}$, and rewriting the
gauge field dependent terms in covariant form, we arrive at the
expression:
\begin{equation}
  \label{eq:seffcs}
 S_{\mathrm{odd}} \;=\; \frac{m \mu}{2 \pi e B}  \, 
\int d^3x \, \epsilon^{\alpha\beta\gamma} \, A_\alpha \partial_\beta A_\gamma \;, 
\end{equation} 
which is a Chern-Simons action, with a coefficient 
\begin{equation}
\kappa \;=\; \frac{\rho e}{ B}
\end{equation}
where we used the fact that $\rho = \frac{m \mu}{2 \pi}$, and we have 
restored the $e$ factors that were absorbed into the gauge field 
definition at the begining of this section. This results
coincides with the one obtained from the commutative field theory
calculation \cite{lf}. We remark that, as in the commutative case, 
the Chern-Simons coefficient will not be renormalized by higher 
order terms in the perturbative expansion in $\tilde A$. Moreover, 
higher order terms in the S-W relation will not contribute to this 
term since they involve higher order powers in $\theta$.

Regarding the general situation, i.e., the case of a general planar
field theory corresponding to charged particles, we note that the
crucial property which is required is the existence of a finite
density (i.e., a finite chemical potential).  The existence of a
finite chemical potential implies that in the evaluation of the
fermionic determinant, the effective action will always have a term
proportional to $\mu$ and to the integral of ${\widehat A}_0$. The
latter, when written in terms of $A_\mu$, is a commutative CS term.

\section {Conclusions}\label{sec:conc}
We have shown that, for a system of non-relativistic fermions in a
magnetic field, the induced CS term for an external gauge field may be
re-derived using the effective noncommutative description corresponding
to the original commutative theory. We insist that the object we
calculate is {\em not\/} the noncommutative induced CS term, but
rather we show how the usual, commutative CS term is indeed captured
by the approximations made to introduce the noncommutative
description.  The corresponding coefficients for the CS terms agree,
and moreover in both the commutative and noncommutative calculations
there are constraints between the density and the magnetic field. In
the former, the reason is that perturbation theory would be
ill-defined for a system with partially filled Landau  levels (degeneracy),
while for the latter the constraint appears in the very derivation of
the noncommutative description (\ref{eq:const}).

\section{Acknowledgements}
This work is supported by CONICET (Argentina), by ANPCyT through grant
No.\ $03-03924$ (AL), and by Fundaci{\'o}n Antorchas (Argentina).
 M. L. C. 
was supported by a Fundaci{\'o}n Antorchas Scholarship. M. L. C.
also acknowledges the kind hospitality of the Particle Physics Group of the Centro
At{\'o}mico Bariloche (Argentina) where part of this work was done.

\end{document}